\documentclass[12pt]{article}
\usepackage{epsfig}
\usepackage{subfigure}
\usepackage{amsmath}
\usepackage{amsfonts}
\usepackage{amssymb}
\newcommand{\be}{\begin{equation}}
\newcommand{\ee}{\end{equation}}
\newcommand{\bea}{\begin{eqnarray}}
\newcommand{\eea}{\end{eqnarray}}
\begin{document}
\begin{titlepage}
\begin{flushright}
hep-ph/0503201
\end{flushright}
\vspace{4\baselineskip}
\begin{center}
{\Large\bf 
Non-thermal Leptogenesis and a Prediction of 
Inflaton Mass in a Supersymmetric SO(10) Model}
\end{center}
\vspace{1cm}
\begin{center}
{\large 
Takeshi Fukuyama
\footnote{E-mail: fukuyama@se.ritsumei.ac.jp}, 
Tatsuru Kikuchi
\footnote{E-mail: rp009979@se.ritsumei.ac.jp}, 
and Toshiyuki Osaka
\footnote{E-mail: rp006002@se.ritsumei.ac.jp}
}
\end{center}
\vspace{0.2cm}
\begin{center}
{\it Department of Physics, Ritsumeikan University, 
Kusatsu, Shiga, 525-8577, Japan}
\medskip
\vskip 10mm
\end{center}
\vskip 10mm
\begin{abstract}
The gravitino problem gives a severe constraint on the thermal 
leptogenesis scenario. This problem leads us to consider some alternatives 
to it if we try to keep the gravitino mass around the weak scale 
$m_{3/2} \sim 100$ GeV. We consider, in this paper, the non-thermal 
leptogenesis scenario in the framework of a minimal supersymmetric 
SO(10) model. Even if we start with the same minimal SO(10) model, we have 
different predictions for low-energy phenomenologies dependent on the types 
of seesaw mechanism. This is the case for leptogenesis: it is shown that 
the type-I see-saw model gives a consistent scenario for the non-thermal 
leptogenesis but not for type-II. The predicted inflaton mass needed to 
produce the observed baryon asymmetry of the universe is found to be 
$M_I \sim 5 \times 10^{11}$ GeV for the reheating temperature 
$T_R = 10^6$ GeV. 

\end{abstract}
\end{titlepage}
\newpage
\section{Introduction}
The supersymmetric (SUSY) grand unified theory (GUT) provides 
an attractive implication for the understandings of the low-energy 
physics. In fact, for instance, the anomaly cancellation between 
the several matter multiplets is automatic in the GUT, since 
the matter multiplets are unified into a few multiplets, and 
the experimental data supports the fact of unification 
of three gauge couplings at the GUT scale, $M_{\rm GUT} = 2 \times 10^{16}$ 
GeV assuming the particle contents of the minimal supersymmetric 
standard model (MSSM) \cite{unification1, unification2}. 
The right-handed neutrino appeared naturally in the SO(10) GUT 
provides a natural explanation of the smallness of the neutrino masses 
through the see-saw mechanism \cite{seesaw}, and also 
the baryon asymmetry of the universe may have its origin in 
the same dimension-five operator relevant to the neutrino masses 
through the leptogenesis scenario \cite{fukugita} that would provide 
a natural explanation for the observed value of the baryon asymmetry 
\cite{data} 
\be
4.9 \times 10^{-11} \ \leq\ Y_B \left(= \frac{n_B}{s} \right) \ \leq\ 
9.9 \times 10^{-11} ~(\mbox{95 \% C.L.})  \;.
\ee
In a series of papers, we have discussed a minimal SO(10) model 
\cite{matsuda1, okada1} and its applications to 
low-energy phenomenologies like neutrino oscillations, 
neutrinoless double beta decay, leptogenesis \cite{okada2}, 
lepton flavour violation \cite{kikuchi1} and proton decay \cite{kikuchi2}. 
(As for the minimal SO(10) models done by the other groups, 
refer \cite{minimal}.) 
In these applications, we achieved rather good coincidences with 
observations. The only exceptional case is leptogenesis where
our theory gives over-production of baryon asymmetry in its naive form 
\cite{okada2}. In order to relieve this pathology, we were forced to make 
the second pair of doublets play some roles unlike the conventional case 
where these doublets are decoupled at the electro-weak scale. 
However, the so called gravitino problem \cite{weinberg} gives a severe 
constraint on the reheating temperature, though this problem has been 
a long history \cite{Khlopov} and there are many studies on this subject 
\cite{Roszkowski}, the recent progress including the hadronic decay processes 
shed a new doubt on the high reheating temperature $T_R$ \cite{kawasaki}. 
If we take it serious and adopt low $T_R$, the thermal leptogenesis becomes 
impossible. This may be good news for us since the minimal SO(10) model in its 
naive form may survive in the new scenario. In the new scenario is not unique 
and we consider in this paper non-thermal leptogenesis, where the inflaton 
decays to the right-handed neutrinos and the successive $B-L$ violating 
decays of the right-handed neutrinos produce the baryon asymmetry. 

\section{Leptogenesis in a minimal SUSY SO(10) model}
Let us first briefly review the conventional leptogenesis scenario 
\cite{fukugita}. 
In the following, our discussion is always based on the effective 
Lagrangian at energies lower than the right-handed neutrino masses 
such that 
\begin{eqnarray}
{\cal L}_{\rm eff}=- 
\int d^2 \theta 
\left(Y_\nu^{ij} N_i^c L_j H_u
+\frac{1}{2} \sum_{i} M_{Ri} N^c_i N_i^c \right)
+ h.c. \; , 
\end{eqnarray} 
where $i,j=1,2,3$ denote the generation indices, 
$Y_\nu$ is the Yukawa coupling, $L$ and $H_u$ are 
the lepton and the Higgs doublets chiral supermultiplets, respectively, 
and $M_{Ri}$ is the lepton-number-violating mass term of the right-handed 
neutrino $N_i$ (we are working on the basis of the right-handed neutrino 
mass eigenstates). The peculiar properties of minimal SO(10) are that we 
can fix $Y_\nu^{ij}$ and $M_{Ri}$ unambiguously 
from the low-energy phenomenologies of quarks and leptons 
\cite{matsuda1, okada1}. 

The lepton asymmetry in the universe is generated by CP-violating 
out-of-equilibrium decay of the heavy neutrinos,  
$N \rightarrow \ell_L H_u^*$ and $N \rightarrow \overline{\ell_L} H_u$. 
The leading contribution is given by the interference between 
the tree level and one-loop level decay amplitudes, 
and the CP-violating parameter is found to be \cite{epsilon}
\begin{eqnarray}
\epsilon = 
\frac{1}{8 \pi (Y_\nu Y_\nu^\dag)_{11}}
\sum_{j=2,3}\mbox{Im} \left[ (Y_\nu Y_\nu^\dag)_{1j}^2 \right]
\left\{ f(M_{Rj}^2/M_{R1}^2)
+ 2 g(M_{Rj}^2/M_{R1}^2) \right\} \; .
\label{epsilon}
\end{eqnarray}
Here $f(x)$ and $g(x)$ correspond to 
the vertex and the wave function corrections, 
\begin{eqnarray}
f(x)&\equiv& \sqrt{x} \left[
1-(1+x)\mbox{ln} \left(\frac{1+x}{x} \right) \right] \;,
\nonumber\\
g(x)&\equiv& \frac{\sqrt{x}}{2(1-x)}   \; ,  
\end{eqnarray}
respectively, and both are reduced to 
$\sim -\frac{1}{2 \sqrt{x}}$ for $ x \gg 1$. 
So in this approximation, $\epsilon$ becomes 
\begin{equation}
\epsilon = - 
\frac{3}{16 \pi (Y_\nu Y_\nu^\dag)_{11}}
\sum_{j=2,3} \mbox{Im} \left[(Y_\nu Y_\nu^\dag)_{1j}^2 \right]
\frac{M_{R1}}{M_{Rj}}\;.
\end{equation}
Using the Type-I see-saw mass of the neutrino, 
$M_\nu = - Y_\nu^T M_R^{-1} Y_\nu \left<H_u \right>^2$, 
$\epsilon$ is further written as \cite{b-y}
\begin{eqnarray}
\epsilon &=& 
\frac{3}{16 \pi} \frac{M_{R1}}{\left<H_u \right>^2}
\frac{\mbox{Im}\left[(Y_\nu M_\nu^* Y_\nu^T)_{11} \right]}
{(Y_\nu Y_\nu^\dag)_{11}}
\nonumber\\
&\equiv& 
\frac{3}{16 \pi}\frac{m_{\nu 3} M_{R1} 
\delta_{\rm eff}}{\left<H_u \right>^2} \;. 
\end{eqnarray}
In the minimal SO(10) model we have the definite form of $Y_\nu$ and 
estimate these values unambiguously.
We have assumed that the lightest $N_1$ decay dominantly contributes 
to the resultant lepton asymmetry. In fact, this is confirmed by numerical 
analysis in the case of hierarchical right-handed neutrino masses 
\cite{plumacher}. Using the above $\epsilon$, 
the generated $Y_{B}$ is described as
\begin{eqnarray}
Y_{B}  \sim  \frac{\epsilon}{g_*}  d \; , 
\end{eqnarray}
where $g_* \sim 100$ is the effective degrees of freedom 
in the universe at $T \sim M_{R1}$, and $ d \leq 1 $ 
is so-called the dilution factor. 
This factor parameterizes how the naively expected value 
$Y_B \sim \epsilon/g_*$ is reduced by washing-out processes. 

We can classify the washing-out processes into two cases with 
and without the external leg of the heavy right-handed neutrinos, 
respectively. The former includes the inverse-decay process 
and the lepton-number-violating scatterings 
mediated by the Higgs boson \cite{luty} such as 
$N + \overline{\ell_L} \leftrightarrow \overline{q_R} + q_L$,  
where $q_L$ and $q_R$ are quark doublet and singlet, respectively. 
The latter case is the one induced 
by the effective dimension five interaction, 
\begin{eqnarray} 
{\cal L}_N = \frac{1}{2} \left(Y_\nu^T M_R^{-1} Y_\nu \right)_{ij}
(L_i H_u)^{T} C^{-1} (L_j H_u) \; , 
\label{4point}
\end{eqnarray}
after integrating out the heavy right-handed neutrinos. 
This term is nothing but the one 
providing the see-saw mechanism \cite{seesaw}. 
The importance of this interaction was discussed in \cite{kolb},  
where the interaction was shown to be necessary 
to avoid the false  generation of the lepton asymmetry 
in thermal equilibrium. While numerical calculations 
\cite{plumacher} \cite{luty} are necessary in order to evaluate 
the dilution factor precisely, $Y_B \sim \epsilon/g_{*}$ roughly 
gives a correct answer, and the washing-out process is mostly not 
so effective. Note that this is the consequence from the current neutrino 
oscillation data as explained in \cite{okada2}. 

The lepton asymmetry parameter $\epsilon$ has been evaluated by 
using the results of the minimal SO(10) model \cite{okada1}, 
and results are listed in the following. 
\begin{center}
\begin{tabular}{|c|c|}
\hline
$\tan \beta $ & $|\epsilon|$ \\
\hline \hline
40 & $ 7.39 \times 10^{-5} $ \\
45 & $ 6.80 \times 10^{-5} $ \\
50 & $ 6.50 \times 10^{-5} $ \\
55 & $ 11.2 \times 10^{-5} $ \\
\hline
\end{tabular}
\end{center}
Unfortunately, the CP-violating parameter $\epsilon$ is too large 
to be consistent with the observed baryon asymmetry. In order to 
circumvent this trouble we made use of another pair of SU(2) doublets 
appearing in the minimal SO(10) model. We solved the Boltzman equation and 
obtained the consistent $Y_B$ \cite{okada2}. However in this case 
we need the extra Higgs other than those in the MSSM, which may raise 
the other problems. So it is deserved to consider an alternative 
solution to this overproduction. 
On the other hand the gravitino problem forces us low reheating temperature 
less than the mass of $M_{R1}$. If we believe it, the above problem becomes 
fake since thermal $N_R$ are not generated in the reheating era. 
So the minimal SO(10) model itself drives us the other approaches 
such as non-thermal leptogenesis scenario \cite{non-thermal} 
or the Affleck-Dine mechanism \cite{Affleck-Dine}. 
In the next section, we discuss on the non-thermal leptogenesis 
scenario in the minimal SO(10) model. 

\section{Non-thermal leptogenesis}
Now we turn to the discussions of the non-thermal leptogenesis scenario 
\cite{non-thermal}. In the non-thermal leptogenesis scenario, 
the right-handed neutrinos are produced through the direct 
non-thermal decay of the inflaton. 

Here we give a concrete model to specify the inflaton. 
We add a singlet chiral supermultiplet which plays a role of inflaton $I$
\footnote{
There is an alternative scenario in which we regard one of the scalar partners 
of the right-handed neutrinos as inflaton \cite{axion}. 
But in this case, we obtained the reheating temperature 
$T_R \sim 4 \times 10^{12}$ [GeV] that is too high for the 
weak scale gravitino. So it is driven by necessity to consider 
the other possibility like a model presented in this paper if 
we keep the gravitino mass at the weal scale. 
}. 
The interaction Lagrangian relevant for the inflaton 
and the right-handed neutrinos is given by 
\be
{\cal L}_I = - \frac{1}{2} \int d^2 \theta
\left(M_I I^2 +\sum_i \lambda_i I N^c_i N^c_i \right)\;.
\label{int}
\ee
When inflaton gets a VEV, it gives rise to the Majorana masses for 
the right-handed neutrinos in addition to the VEV of 
$({\bf 10,1,3})$ in $\overline{\bf 126}$ 
under $SU(4)_{PS} \times SU(2)_L \times SU(2)_R$ \cite{matsuda1, okada1}. 
However, the VEV $\left<I \right>$ is posted around the GUT scale and 
$\lambda_i$ is found to be $10^{-8}$ later, and this contribution gives 
a tiny correction to $M_R$. Also the first term in Eq.~(\ref{int}) 
dominates over the second, and is reduced to the chaotic inflationary 
model \cite{chaotic}. 

\begin{figure}[h]
\begin{center}
\subfigure[The lepton asymmetry parameter $\epsilon$ has been taken from 
\cite{okada1}]
{\includegraphics*[width=.9\linewidth]{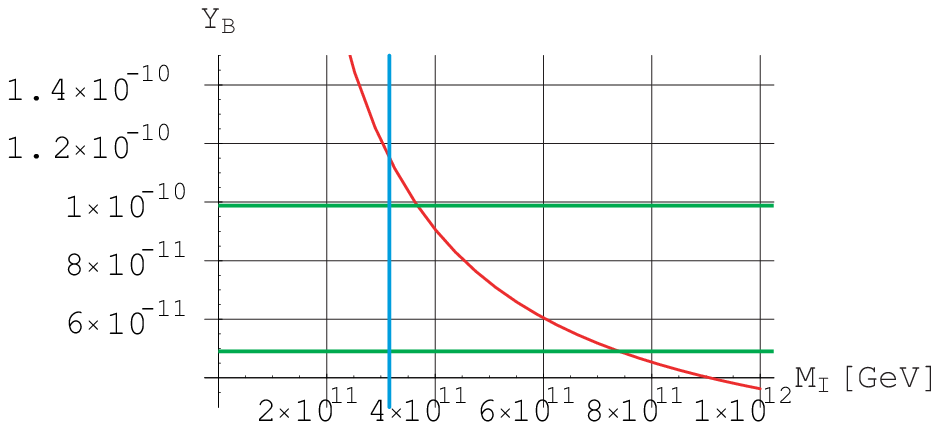}
\label{Fig1a}}
\subfigure[The lepton asymmetry parameter $\epsilon$ has been taken from 
\cite{mohapatra}]
{\includegraphics*[width=.7\linewidth]{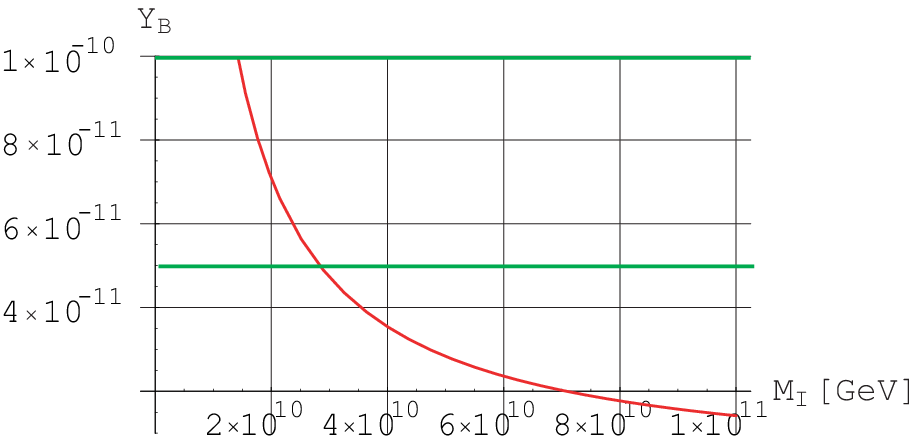}
\label{Fig1b}}
\caption{The predicted baryon asymmetry of the universe $Y_B = n_B/s$ 
as a function of the inflaton mass $M_I ~\mbox{[GeV]}$ with the reheating 
temperature $T_R = 10^6 ~{\rm GeV}$ and ${\rm BR} = 1$. In Fig.~\ref{Fig1a}, 
the vertical blue line represents the kinematical cut for the inflaton to 
have enough energy to decay, $E \geq 2 M_{R1}$, {\it i.e.}, 
the right hand side of this line is allowed from the kinematics. 
Two horizontal green lines represent the upper and the lower bounds on 
the observed value of the baryon asymmetry at 95 \% C.L. }
\label{Fig1}
\end{center}
\end{figure}

In such a superpotential, the inflaton decay rate is given by
\be
\Gamma(I \to N_i N_i) \simeq \frac{|\lambda_i|^2}{4 \pi} M_I \;.
\ee
Then the consequently produced reheating temperature is obtained by
\be
T_R \ =\ \left(\frac{45}{2 \pi^2 g_*} \right)^{1/4}
(\Gamma M_P)^{1/2} \;.
\ee
If the inflaton dominantly couples to $N$, the branching ratio of this 
decay process is, of course, ${\rm BR} \sim 1$. 
Then the produced baryon asymmetry of the universe can be calculated 
by using the following formula, 
\bea
\left(\frac{n_B}{s} \right)
&=& - 0.35 \times \left(\frac{n_{N_1}}{s} \right)
\times \left(\frac{n_L}{n_{N_1}} \right)
\nonumber\\
&=& - 0.35 \times \frac{3}{2} ~{\rm BR}(I \to N_1 N_1)
\left(\frac{T_R}{M_I} \right) \times \epsilon\;.
\label{YB}
\eea
With the hierarchical mass spectra for the right-handed neutrinos, 
it can be approximated as 
\be
\left(\frac{n_B}{s} \right)
= - 1.95 \times 10^{-10} \times {\rm BR} \times
\left(\frac{T_R}{10^6~{\rm GeV}} \right)
\left(\frac{M_{R1}}{M_I}\right) 
\left(\frac{m_{\nu 3}}{0.065~{\rm eV}}\right) \times 
\delta_{\rm eff} \;,
\ee
where $\delta_{\rm eff} \equiv 
\mbox{Im}\left[(Y_\nu M_\nu^* Y_\nu^T)_{11} \right]/
\left[m_{\nu 3} (Y_\nu Y_\nu^\dag)_{11} \right]$ 
denotes the effective value of the CP violating phase parameter 
relevant for the leptogenesis and it can be estimated as 
$\delta_{\rm eff} = - 0.166$ in our model. 
As it can easily be seen that it is possible to produce the baryon asymmetry 
of the universe by using the reheating temperature as low as, 
$T_R \lesssim 10^6 ~{\rm GeV}$. Hence, a very wide range of the gravitino mass 
can be allowed, $m_{3/2} \gtrsim 10^{6} ~{\rm MeV}$. 
The result of the detailed numerical calculation based on Eq.~(\ref{YB}) 
is shown in Fig.~\ref{Fig1}. 
As shown in Fig.~\ref{Fig1a} that the predicted inflaton mass is 
heavier than the lightest right-handed neutrino 
($M_{R1} = 1.6 \times 10^{11}$ GeV) in our model \cite{okada1}. 
Hence the non-thermal leptogenesis is well workable. But in model 
\cite{mohapatra}, you can see from Fig.~\ref{Fig1b} that 
the calculated inflaton mass is lighter than the lightest right-handed 
neutrino mass ($M_{R1} = 2.7 \times 10^{13}$ GeV), and the non-thermal 
leptogenesis scenario is prohibited by the kinematics. 
We hasten to add that this conclusion is valid under the non-thermal 
leptogenesis under the gravity mediated SUSY breaking scenario. 

It can be read off from Fig.~\ref{Fig1a} that the observed value of 
the baryon asymmetry leads to the inflaton mass around 
$M_I \sim 5 \times 10^{11}$ GeV. 
This corresponds to the coupling constant of the inflaton to 
the right-handed neutrinos as $\lambda_i \sim 10^{-8}$. 
Such a small coupling indicates that the model can naturally fit into
the chaotic inflationary model \cite{chaotic} 
based on a minimal supersymmetric SO(10) model. 

\section{Summary}
In confronting with the recent progress on the gravitino problem 
\cite{kawasaki}, the usual thermal leptogenesis scenario encounters 
some problems. In this paper, we have explored the non-thermal leptogenesis 
as an alternative scenario to the thermal one. We have estimated 
the baryon asymmetry of the universe based on a minimal supersymmetric 
SO(10) model with type-I see-saw mechanism. The result of our analysis 
shows that the non-thermal scenario well works within the minimal SO(10) 
model, and we have found the inflaton mass around $M_I \sim 5 \times 10^{11}$ 
GeV gives the observed value of the baryon asymmetry of the universe. 
In this analysis, we have used the reheating temperature $T_R = 10^6$ GeV 
which was chosen so as to realize the weak scale gravitino mass 
$m_{3/2} \sim 100$ GeV without causing the gravitino problem. 
Even if these values are relaxed by one order of magnitude 
($m_{3/2} \lesssim 10~\mbox{TeV},~T_R = 10^7$ GeV), 
the result is still valid. 

In this paper, we have assumed that full thermalization occurs soon after 
the inflation. Although there is a discussion on this point \cite{Mazumdar2}, 
the main motivation of this paper is to give a quantitative estimation 
for the non-thermal leptogenesis scenario in the minimal SO(10) model. 
Therefore, we leave the consideration about thermalization processes 
for future study.

\section*{Acknowledgments}
The work of T.F. is supported in part by the Grant-in-Aid for 
Scientific Research from the Ministry of Education, Science and Culture 
of Japan (\#16540269). He is also grateful to Professors D. Chang and 
K. Cheung for their hospitality at NCTS. The work of T.K. are supported 
by the Research Fellowship of the Japan Society for the Promotion 
of Science (\#7336).

\end{document}